\documentclass[twocolumn,showpacs,preprintnumbers,amsmath,amssymb]{revtex4}

\usepackage{graphicx,dcolumn,bm,times}

\begin{document}

\title{Geographical effects on epidemic spreading in scale-free networks}

\author{Xin-Jian Xu,$^{1}$ Wen-Xu Wang,$^{1,2}$ Tao Zhou,$^{1,2}$ and Guanrong Chen$^{1,}$\footnote{Electronic address: gchen@ee.cityu.edu.hk}}

\address{$^{1}$Department of Electronic Engineering, City
University of Hong Kong, 83 Tat Chee Avenue, Kowloon, Hong Kong SAR, China\\
$^{2}$Nonlinear Science Center and Department of Modern Physics,
University of Science and Technology of China, Hefei Anhui 230026,
China}

\date{\today}

\begin{abstract}
Many real networks are embedded in a metric space: the
interactions among individuals depend on their spatial distances
and usually take place among their nearest neighbors. In this
paper, we introduce a modified susceptible-infected-susceptible
(SIS) model to study geographical effects on the spread of
diseases by assuming that the probability of a healthy individual
infected by an infectious one is inversely proportional to the
Euclidean distance between them. It is found that geography plays
a more important role than hubs in disease spreading: the more
geographically constrained the network is, the more highly the
epidemic prevails.
\end{abstract}


\maketitle

Accurately modelling epidemic spreading is an important topic in
understanding the impact of diseases and the development of
effective strategies for their control and containment \cite
{Anderson}. The classical mathematical approach for describing
disease spreading either ignores the population structure or
treats population as distributed in a uniform medium. However, it
has been argued in the past few years that many social,
biological, and communication systems possess two universal
characters, the small-world effect \cite {Watts} and the
scale-free property \cite {Barabasi}, which can be described by
complex networks whose nodes represent individuals and links
represent the interactions among them \cite {Albert}. In view of
the wide occurrence of complex networks in nature, it is
interesting to study the effects of topological structures on the
dynamics of epidemic spreading. Pioneering works \cite {Moore,
May, Kuperman, Pastor_1, Moreno_1} have given some valuable
insights: for homogeneous networks (e.g., regular, random, and
small-world networks), there are critical thresholds of the
spreading rate below which infectious diseases will eventually die
out; on the contrary, even infections with low spreading rates
will prevail over the entire population in heterogeneous networks
(e.g., scale-free networks). This radically changes many
conclusions drawn from classic epidemic modelling. Furthermore, it
has been observed that the heterogeneity of a population network
in which the disease spreads may have noticeable effects on the
evolution of the epidemic as well as the corresponding
immunization strategies \cite {Pastor_1, Moreno_1, Cohen_2,
Barthelemy_1}.

In many real networks, however, individuals are often embedded in
a Euclidean geographical space and the interactions among them
usually depend on their spatial distances and take place among
their nearest neighbors \cite {Durrett, Yook, Nemeth, Gastner}.
For instance, the number of long-range links and the number of
edges connected to a single node are limited by the spatial
embedding, particularly in planar networks. Also, it has been
proved that the characteristic distance plays a crucial role in
the dynamics taking place on these networks \cite {Manna,
Rozenfeld, ben-Avraham, Warren, Huang, Xie}. Thus, it is natural
to study complex networks with geographical properties. Rozenfeld
\emph{et al.} considered that the spatial distance can affect the
connection between nodes and proposed a lattice-embedded
scale-free network (LESFN) model \cite {Rozenfeld}. Based on a
natural principle of minimizing the total length of links in the
system, a scale-free network can be embedded in a Euclidean space.
Since distributions of individuals in social networks always
depend on their spatial locations, the study of the influence of
geographical structures on dynamical processes is of great
importance.

\begin{figure}
\includegraphics[width=\columnwidth]{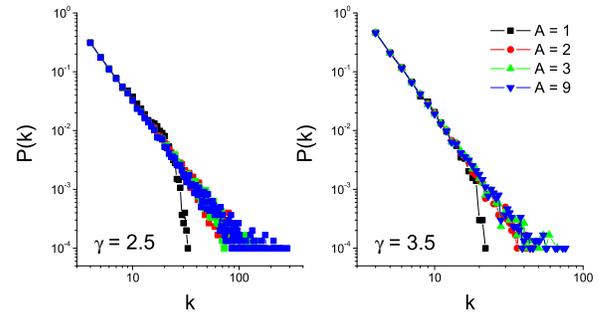}

\caption{(color online) Degree distribution of the LESFN model
with $N=10000$ for $\gamma=2.5$ and $3.5$. The territory parameter
$A$ controls the influence of the geographical distance on the
network structure.}\label{pk}
\end{figure}

In this paper, we present a modified
Susceptible-Infected-Susceptible (SIS) model on the LESFN to
investigate how the geographical structure affects the dynamical
process of epidemic spreading. Here, we assume that the time
scales governing the dynamics is much smaller than those
characterizing the network evolvement. Thus, the static network is
suitable to use for discussing the problem under investigation. In
contrast to the assumption that the infection probability is
identical across successive contacts, we define the probability of
a healthy individual $i$ infected by an infectious one $j$ to be
inversely proportional to the Euclidean distance between them.
Based on computer simulations, we found that when the network
connectivity is less local, it will be more robust to disease
spreading, regardless of the heterogeneous distribution of nodes.

The LESFN is generated as follows \cite {Rozenfeld, ben-Avraham}:
(i) a lattice of size $N=L \times L$ with periodic boundary
conditions is assumed, upon which the network will be embedded;
(ii) for each site of the lattice, a preset degree $k$ is assigned
taken from a scale-free distribution, $P(k) \sim k^{-\gamma}$,
$m<k<K$; (iii) a node (say $i$, with degree $k_{i}$) is picked
randomly and connected to its closest neighbors, until its degree
quotum $k_{i}$ is realized or until all sites up to a distance
have been explored
\begin{equation}
d(k_{i})=A\sqrt{k_{i}}, \label{distance}
\end{equation}
Duplicate connections are prohibited. Here, $d(k_{i})$ is the
spatial distance on a Euclidean plane denoting the characteristic
radius of the region that node $i$ can almost freely to reach the
others; (iv) this process is repeated throughout all the sites on
the lattice. Following this method, networks with $\gamma
> 2$ can be successfully embedded up to a (Euclidean) distance
$d(k)$ which can be made as large as desired upon the change of
the territory parameter $A$. The model turns out to be a randomly
connected scale-free network when $A \rightarrow \infty$ \cite
{Newman_2}. Typical networks with $\gamma=2.5$ and $3.5$ resulting
from the embedding method are illustrated in Fig. \ref{pk}. In the
case of $N=10000$, the power-low degree distributions of the
LESFNs achieve their natural cutoff lengths for $A=2$, $3$ and
$9$, while they end at some finite-size cutoff lengths for $A=1$.

In order to study geographical effects on the spread of diseases,
we introduce a modified SIS model. In this model, an individual is
described by a single dynamical variable adopting one of the two
stages: \emph{susceptible} and \emph{infected}. Considering the
geography, we assume that the probability of a healthy individual
$i$ infected by an infectious one $j$ is inversely proportional to
the Euclidean distance between them, defined by
\begin{equation}
\lambda_{ij}=\frac{1}{d_{ij}^{\alpha}}, \label{spreadrate}
\end{equation}
where $\alpha$ is a tunable parameter. This is motivated by the
following idea: human beings are located in territories and they
interact more frequently with their nearest neighbors than those
far away. The transmission of a disease is described in an
effective way with the following rules: a susceptible individual
at time $t$ will pass to the infected state with the rate
$\lambda$ at time $t+1$ if he is connected to infected
individuals. Infected individuals at time $t$ will pass to the
susceptible state again with the unite rate at time $t+1$.
Individuals run randomly through the cycle, \emph{susceptible}
$\rightarrow$ \emph{infected} $\rightarrow$ \emph{susceptible}.

\begin{figure}
\includegraphics[width=\columnwidth]{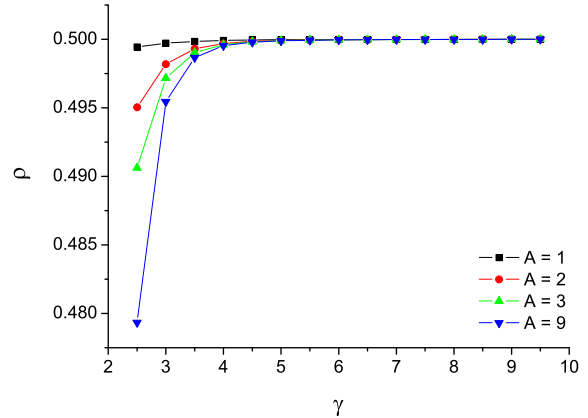}

\caption{(color online) Density of infected individuals in the
stationary state $\rho$ vs. the scale-free degree exponent
$\gamma$ for the LESFNs with different values of $A$. The results
are obtained for $\alpha=2$ and on networks of size
$N=10000$.}\label{rhogamma}
\end{figure}

In the present work, we have performed Monte-Carlo (MC)
simulations of the model with synchronously updating on the
network. Initially, the number of infected nodes is $1 \%$ of the
size of the network. The total sampling times are $10000$ (MC time
steps). After appropriate relaxation times, the systems is
stabilized to a steady state. Simulations were implemented on the
network model averaging over $500$ different realizations. Given a
network, an important observable is the prevalence $\rho$, which
is the time average of the fraction of infected individuals in the
steady state (averaging over $1000$ time steps in this context).

Figure \ref{rhogamma} shows the persistence of infected
individuals $\rho$ versus the scale-free degree exponent $\gamma$
for the LESFNs with different values of $A$ when $\alpha$ is fixed
at $2$. As $\gamma$ increases, all the curves approach to an
asymptotic value of $\rho=0.5$, independent of the geography of
networks. The larger the parameter $A$, the quicker the prevalence
$\rho$ is close to the asymptotic value. This implies that the
scale-free degree exponent has a slight influence on the spread of
diseases when networks are more geographical constrained (smaller
$A$) in comparison with the case of more scale-free region (larger
$A$). It has been suggested \cite {Vazquez} that there is a
threshold $\gamma_{c}=3$ which separates the two different
dynamical behaviors of disease spreading, so we will focus on the
values of $\gamma$ at $2.5$ and $3.5$, respectively, to study two
typical cases.

\begin{figure}
\includegraphics[width=\columnwidth]{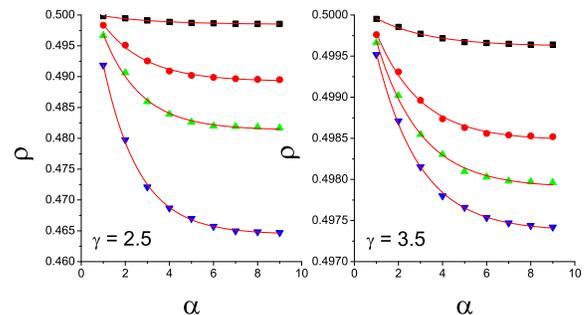}

\caption{(color online) Density of infected individuals $\rho$ vs.
the tunable parameter $\alpha$ for the LESFNs with different
values of $A$: $A=1$ (squares), $2$ (circles), $3$ (up triangles),
and $9$ (down triangles), respectively. The network size is
$N=10000$.}\label{rhoalpha}
\end{figure}

\begin{figure}
\includegraphics[width=\columnwidth]{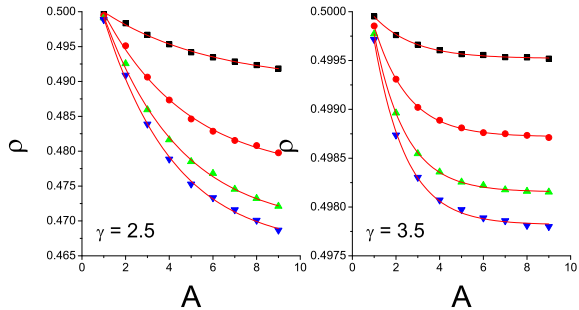}

\caption{(color online) Density of infected individuals $\rho$ vs.
the territory parameter $A$ of the modified SIS model with
different values $\alpha$: $\alpha=1$ (squares), $2$ (circles),
$3$ (up triangles), and $4$ (down triangles), respectively. The
network size is $N=10000$.}\label{rhoa}
\end{figure}

In Fig. \ref{rhoalpha}, we plot the densities of infected
individuals $\rho$ versus the tunable parameter $\alpha$ for the
LESFNs with $\gamma=2.5$ and $3.5$, respectively. As $\alpha$
becomes larger, the prevalence $\rho$ decreases, and all the
curves approach to stable values finally. The solid lines fits to
the form $\rho = B_{0} + B e^{-\alpha/\alpha_{0}}$, implying that
there is a relation of the first-order exponential decay between
$\rho$ and $\alpha$. According to the definition of the spreading
probability of our model (see Eq. (\ref{spreadrate})), one can
easily find that the larger the parameter $\alpha$, the smaller
the spreading rate $\lambda$ is. This results in a small fraction
of infected nodes in the network. In order to understand how the
geographical structure affects the epidemic dynamics, we plot the
prevalence $\rho$ as a function of the territory parameter $A$ in
Fig. \ref{rhoa}. It shows that as $A$ increases, the density
$\rho$ deceases. In other words, when networks are more
geographically constrained, i.e., more locally interconnected,
they tend to have larger epidemic prevalence. This is different
from the results observed on Barab\'{a}si-Albert scale-free
networks, where nodes with large degrees (called \lq\lq
hubs\rq\rq) accelerate the spreading process and induce
significant epidemic prevalence. The solid line fits to the form
$\rho = C_{0} + C e^{-A/A_{0}}$. Similar to Fig. \ref{rhoalpha},
there is also a relation of the first-order exponential decay
between $\rho$ and $A$.

\begin{figure}
\includegraphics[width=\columnwidth]{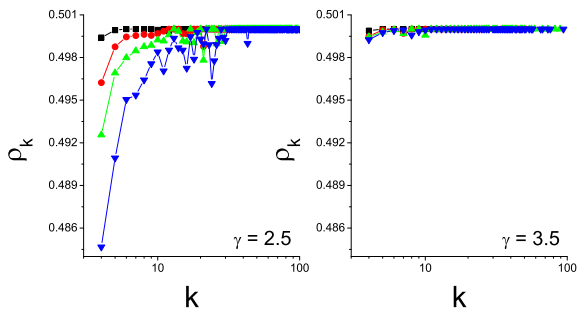}

\caption{(color online) The density $\rho_{k}$, defined as the
fraction of nodes with connectivity $k$ that are infected, in
LESFNs of size $N=10000$ and the territory parameters $A=1$
(squares), $2$ (circles), $3$ (up triangles), and $9$ (down
triangles), respectively.}\label{rhok}
\end{figure}

We also provide an illustration for the behavior of the
probability $\rho_{k}$ that a node with given connectivity $k$ is
infected. The range of an edge is the length of the shortest paths
between the nodes it connected in the absence of itself \cite
{Watts_1, Pandit}. If an edge's range is $l$, the shortest cycle
it lies on is of length $l+1$. Thus the distribution of range in a
network sketches the distribution of shortest cycles. It has been
demonstrated numerically that when the spatial constraint is
stronger, the LESFN has more small-order cycles \cite {Huang}. In
this case, the nodes are more likely to meet which speeds disease
spreading. As shown in Fig. \ref{rhok}, in the case of
$\gamma=2.5$, there is a heterogeneous behavior of $\rho_{k}$,
especially for $A=9$, i.e., the higher a node degree, the larger
the probability $\rho_{k}$. However, when a node's degree is
larger than a certain value, this feature vanishes. This implies
that even a geographical network tends to a scale-free random
graph, where the hub effect on the spreading dynamics is still
limited, i.e., a node's potential infectivity is not strictly
equal to its degree due to geographical effects. This effect is
more stronger for $\gamma=3.5$, that is, all the curves are nearly
linearly independent of the value of the territory parameter.

In conclusion, we have studied geographical effects on the
spreading phenomena in lattice-embedded scale-free networks, in
which a territory parameter $A$ controls the influence of the
geography on the network structure and therefore on the epidemic
dynamics. We studied the modified SIS model in which the
probability of a healthy individual infected by an infectious one
is inversely proportional to the Euclidean distance between them.
Our main finding is that when the network is more geographically
constrained, i.e., with heavier local connections, the epidemic
prevalence will be more significant. This indicates that networks
with more local connections have a higher risk to disease
spreading. On the contrary, while the network is more scale-free,
it will be more robust to disease spreading, regardless of the
heterogeneous connectivity of the network.

G. Chen acknowledges the Hong Kong Research Grants Council for the
CERG Grant CityU 1114/05E. T. Zhou acknowledges the Natural
Science Foundation of China for the Grant No. 70471033.

\end{document}